\documentclass[twocolumn,showpacs,preprintnumbers,amsmath,amssymb]{revtex4}

\usepackage{graphicx}
\usepackage{dcolumn}
\usepackage{bm}
\usepackage[dvips]{color}
\usepackage{amsmath}
\usepackage{amssymb}
\usepackage{mathrsfs}

\newcommand{\Lin}{\mathcal{L}}
\newcommand{\Ker}{\mathcal{K}}

\newcommand{\w}{\omega}

\newcommand{\tr}{{\rm Tr}}

\newcommand{\dd}{{\rm d}}

\begin{document}

\title{Spectral Analysis and Identification of Noises in Quantum Systems}

\author{R.~B. Wu$^{1,2,3}$}
\author{T.~F. Li$^{3,5,6}$}
\author{A.~G. Kofman$^{2,7}$}
\author{J. Zhang$^{1,2,3}$}
\author{Yu-xi Liu$^{2,3,5}$}
\author{Yu.~A. Pashkin$^{4,6}$}
\author{J.-S. Tsai$^{6}$}
\author{Franco Nori$^{2,7}$}

\affiliation{$^1$Department of Automation, Tsinghua University,
Beijing, 100084, China}
\affiliation{$^2$Advanced Science Institute, RIKEN, Saitama, 351-0198, Japan}
\affiliation{$^3$Center for Quantum Information Science and Technology, TNlist, Beijing, 100084, China}
\affiliation{$^4$Physics Department, Lancaster University, Lancaster LA1 4YB, United Kingdom}
\affiliation{$^5$Institute of Micro-Nano Electronics, Tsinghua University,
Beijing, 100084, China}
\affiliation{$^6$NEC Smart Energy Research Laboratories, Tsukuba, Ibaraki 305-8501, Japan}
\affiliation{$^7$Physics Department, The University of Michigan, Ann Arbor, Michigan
48109-1120, USA}

\begin{abstract}
In quantum information processing, knowledge of the noise in the system is crucial for high-precision manipulation and tomography of coherent quantum operations. Existing strategies for identifying this noise require the use of additional quantum devices or control pulses. We present a noise-identification method directly based on the system's non-Markovian response of an ensemble measurement to the noise. The noise spectrum is identified by reversing the response relationship in the frequency domain. For illustration, the method is applied to superconducting charge qubits, but it is equally applicable to any type of qubits. We find that the identification strategy recovers the well-known Fermi's golden rule under the lowest-order perturbation approximation, which corresponds to the Markovian limit when the measurement time is much longer than the noise correlation time. Beyond such approximation, it is possible to further improve the precision at the so-called optimal point by incorporating the transient response data in the non-Markovian regime. This method is verified with experimental data from coherent oscillations in a superconducting charge qubit.
\end{abstract}

\pacs{03.67.-a, 42.50.Lc, 85.25.Dq}

\maketitle

\section{Introduction}

Artificial atoms fabricated with solid-state devices (e.g., superconducting qubits, quantum dots, and NV centers) are promising for future quantum information processors (e.g., \cite{You2005,Buluta2009,Clerk2010,Buluta2011,Makhlin2001} and references therein). To construct more stable and robust quantum circuits, much effort is required to overcome complex decoherence brought by their ``dirty" environment. Higher performance can only be achieved by taking control actions (e.g., the dynamical decoupling \cite{Viola1998,Uhrig2007,Du2009,Khodjasteh2010} or environment modulation \cite{Kofman2001,Gordon2007a,Kofman2004,Gordon2009}) to fight against decoherence. To characterize and control quantum systems, it is crucial to obtain a dynamical model of the quantum system according to the measurement result of the system. For closed systems, this calls for the identification of the Hamiltonian \cite{Bonnabel2009,ReydeCastro2010,Geremia2002}. Otherwise, when the system is open, the dissipation effect also needs to be specified corresponding to the noises coupled to the system. Hence, the information acquisition of the noises is critical for modeling open quantum system dynamics. The knowledge of noise is also useful for improving the fidelity of state and process tomography from a series of designed quantum measurements \cite{Chuang1997,Liu2005,Childs2001,Riebe2006,Neeley2008}. Since the number of required measurements for process tomography increases quickly with the system dimensionality, a good model for the dynamics of the measured system will help reduce the measurement cost based on the prediction of the actual process matrix.

\begin{figure}
  \includegraphics[width=3.4 in]{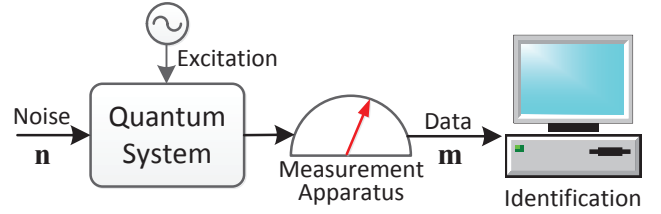}
  \caption{(color online) Schematics for a general identification procedure. The quantum dynamics is properly excited so that the noise $\mathbf{n}$ can have a simple relation with the measurement data $\mathbf{m}$, which is then reversed to numerically identify the noise properties.}\label{schematics}
\end{figure}

Physically, a stationary noise in a quantum system is mainly characterized by its correlation spectrum \cite{Kubo1978}. In many solid-state systems, the spectrum has nontrivial structures (e.g., Lorentzian shape, Ohmnic, 1/f noise \cite{Breuer2002}), which corresponds to colored noise. Figure \ref{schematics} shows the general procedure for identifying an unknown noise $``\mathbf{n}"$ acting on a quantum system. The expectation value $\mathbf{m}$ of some observable is measured. If one knows how the noise $\mathbf{n}$ affects $\mathbf{m}$ via a functional mapping $\mathbf{m}=\mathscr{S}(\mathbf{n})$ [e.g., Eq.~\eqref{Optimal Point} in Sec.\ref{Sec:Qubit}], then this mapping can be reversed {(if reversible)} to identify the noise from the measurement data, i.e., $\mathbf{n}=\mathscr{S}^{-1}(\mathbf{m})$ [e.g., Eq.~\eqref{Identification 1} in Sec.\ref{Sec:Qubit}]. Hence, the identification precision largely depends on how well the model $\mathbf{m}=\mathscr{S}(\mathbf{n})$ describes the real noise-drive quantum dynamics.

Generally, the noise causes a random frequency shift and dissipation effects in the system dynamics, which can be described by a generalized master equation \cite{Kubo1978,Kofman2004,Clerk2010} or a spectral overlap function with the noise correlation spectrum \cite{Kofman2000,Kofman2004}. Such models were used to derive \cite{Schoelkopf2003,Almog2011} the widely used identification strategy based on Fermi's golden rule, which claims that the state transition rate of a two-level system is proportional to the intensity of the noise spectrum at the resonating frequency. This makes it possible to experimentally analyze the noise by using a frequency-tunable qubit \cite{Gavish2000,Aguado2000,Schoelkopf2003,Clerk2010,Almog2011}. Moreover, in Refs.\cite{Yuge2011,Alvarez2011}, the dephasing noise spectrum is identified from the direct measurement of the asymptotic decay of the nuclear spin qubit in NMR experiments. The strategy requires designed $\pi$-pulse sequences to cancel the dominant influence of the lower-frequency component of the noise spectrum \cite{Savel¡¯ev2010,Satanin2012}

The above identification schemes essentially exploit only the system's long-time (compared to the noise correlation time) response to the noise \cite{Schoelkopf2003}, while the non-Markovian transient dynamics is not considered. In this paper, we introduce a new identification method that incorporates the non-Markovian noise-driven dynamics. The model adopted for identification is based on the Born approximation, which has been used to compute non-Markovian dynamics driven by known colored noises \cite{Liu2001,DiVincenzo2005,Burkard2009,Marcos2011}. Under the Laplace transform, the model appears in a purely algebraic form that will simplify the analysis for the purpose of noise identification.

This paper is organized as follows. In Section \ref{Sec:modeling}, a frequency-domain model is derived for analyzing the response of any ensemble measurement to the noise in a quantum system. In Section \ref{Sec:Qubit}, the model is applied to a qubit system and is shown to be consistent with Fermi's golden rule. Then, from the model we propose a new identification strategy at the optimal point and demonstrate it with experimental data. Finally, Section \ref{Sec:Conclusion} presents our conclusions.

\section{Frequency-domain Modeling of Open Quantum Systems}\label{Sec:modeling}

In this section, we will introduce the generalized master equation based on the Born approximation, and Laplace transform the equation to the frequency-domain for the following identification analysis. Although such model has been obtained in different ways in the literature \cite{DiVincenzo2005,Burkard2009,Marcos2011}, we will express it in a new compact form to highlight the influences of the noise on the qubit dynamics. This will facilitate the following study on identifying the noise spectrum from the measurement data of a physical observable.

Consider an $N$-level quantum system that is coupled to some unknown environment. Suppose that the total Hamiltonian is
$$ H_{\rm tot}= H_0+H_1\otimes C(t),$$
where $H_0$ is the internal Hamiltonian of the system. The environment operator $C(t)$ coupled to the system's observable $H_1$ is responsible for the noise. For convenience, it has been rotated to the interaction picture.
Assume that the environment is initially at an equilibrium state $\rho_B^0$ and the coupling strength between the system and the environment is so weak that the joint evolution with the system hardly alters the state of the environment. This qualifies the Born approximation \cite{DiVincenzo2005}, under which the net system evolution (after averaging out the environment) can be described by the following generalized master equation (with $\hbar=1$) \cite{Kofman1994,Kofman2004,DiVincenzo2005,Gordon2007a,Burkard2009,Marcos2011}:
\begin{eqnarray}\label{rho-ME}
\nonumber    \dot \rho(t)\!\! &=&\!\! -i\left[ H_0,\rho(t)\right] - \int_0^t\!\!\dd t'\!\left\{ \Phi(t,t')\left[H_1,e^{-i(t-t')H_0}\times \right.\right.\\
    && \quad \left.\left.H_1\rho(t')e^{i(t-t')H_0}\right]+{\rm H.c.}\right\},
\end{eqnarray}
where $\rho(t)$ is the density matrix of the system.

The integral term in the above equation captures the noise-driven quantum dynamics, where the environmental noise enters through the temporal noise correlation function
\begin{equation}\label{Eq: def of Phi}
    \Phi(t,t')=\tr\left[C(t)C(t')\rho_{B}^0\right]=\Gamma(t,t')+i\Omega(t,t'),
\end{equation}
where $\Gamma(t,t')$ and $\Omega(t,t')$ are the real and imaginary parts of $\Phi(t,t')$, respectively.
Because the environment is initially prepared at an equilibrium state $\rho_B^0$, the correlation function only depends on the time difference, $t-t'$. Thus, in the following, $\Phi(\cdot)$ [as well as $\Gamma(\cdot)$ and $\Omega(\cdot)$] will be always taken as a single-variable function of $\tau=t-t'$. The system behaves in a non-Markovian manner when the time is comparable with the noise correlation time.

Here, $\Gamma(t,t')$ and $\Omega(t,t')$ in the correlation function $\Phi(t,t')$ correspond to the commutative and non-commutative parts of $C(t)$, respectively, as follows \cite{Clerk2010}:
\begin{eqnarray*}
\Gamma(\tau)&=&\frac{1}{2}\tr\left\{\left[C(\tau)C(0)+C(0)C(\tau)\right]\rho_{B}^0\right\}, \\
\Omega(\tau)&=&\frac{1}{2i}\tr\left\{\left[C(\tau)C(0)-C(0)C(\tau)\right]\rho_{B}^0\right\}.
\end{eqnarray*}
In particular, when the environment is at zero temperature, $\Gamma(t)$ corresponds to the average dissipation rate, while $\Omega(t)$ corresponds to the Lamb shift.


To facilitate the following analysis, we first transform Eq.~(\ref{rho-ME}) for the density matrix $\rho(t)$ into a vector form. Choose $\{M_0=N^{-1}I_N,M_1,\cdots,M_{N^2-1}\}$ as an orthonormal basis of the space of $N\times N$ Hermitian matrices, {where each $M_i$ is Hermitian and $\tr(M_iM_j)=\delta_{ij}$ for any $0\leq i\leq j\leq N^2-1$}. Then, the density matrix $\rho(t)$ can be spanned as
$$\rho(t)=v_0(t)M_0+\cdots+v_{N^2-1}(t)M_{N^2-1},$$
where $v_i(t)=\tr\left[\rho(t)M_i\right]$. The real vector $v(t)=[v_0(t),\cdots,v_{N^2-1}(t)]^T$ is called the augmented Bloch representation of $\rho(t)$ (see Section \ref{Sec:Qubit} for the example of two-level systems). In this way, Eq.~\eqref{rho-ME} can be translated into the following form:
\begin{eqnarray}\label{x-ME}
\nonumber    \dot {v}(t)\!\!&=&\!\!\Lin_0 v(t)+\Lin_1\int_0^t\!\!\dd t'\left\{ \Gamma(t-t') e^{(t-t')\Lin_0}\Lin_1 \right.\\
    && \quad \left.+\Omega(t-t') e^{(t-t')\Lin_0}\Lin_1^+\right\} v(t'),
\end{eqnarray}
where $\Lin_{k}$ and $\Lin_{k}^+$, $k=0,1$, are the matrix representations of the commutator operation $[-iH_k,\cdot]$ and the anti-commutator operation $\{H_k,\cdot\}$, respectively.

Eq.~\eqref{x-ME} can be naturally Laplace transformed to the frequency domain as only a linear term and a convolution term of $v(t)$ are involved on the right hand side. This gives
\begin{eqnarray}\label{s-ME}
 sv(s)-v^0\!\!&=&\!\!\Lin_0  v(s)+ \Lin_1\mathcal{K}(s) v(s),
\end{eqnarray}
where $v^0$ is the Bloch vector corresponding to the initial density matrix. The matrix $s$-function
\begin{eqnarray}
\nonumber \mathcal{K}(s)&=&\mathscr{L}\!\left[\Gamma(\tau)e^{\tau\Lin_0 }\right]\Lin_1+ \mathscr{L}\!\left[\Omega(\tau)e^{\tau\Lin_0 }\right]\Lin_1^+ \\
\label{Eq:K(s)} &=& \Gamma(s\mathbb{I}-\Lin_0)\Lin_1+ \Omega(s\mathbb{I}-\Lin_0)\Lin_1^+
\end{eqnarray}
characterizes the action of the noise on the system evolution, where {$\mathscr{L}[\cdot]$ denotes the Laplace transform} and $\mathbb{I}$ is the identity operator on the space of Bloch vectors. Here the property $\mathscr{L}[c(t)e^{s_0t}]=c(s-s_0)$ of the Laplace transform of a scalar function $c(t)$ has been extended to matrix-valued functions of $\Gamma(s\mathbb{I}-\Lin_0)$ and $\Omega(s\mathbb{I}-\Lin_0)$ (see Appendix A for details).

Let us now denote by the matrix $s$-function $\mathcal{R}(s)=(s \mathbb{I}-\Lin_0)^{-1}$ the resolvent operator for the unperturbed system evolution. One can immediately derive from Eq.~\eqref{s-ME} the following frequency domain model:
\begin{equation}\label{s-domain}
 v(s)=\left[\mathcal{R}^{-1}(s)-\Lin_1\mathcal{K}(s)\right]^{-1}v^0,
\end{equation}
as well as that of the time derivative $\gamma(t)=\dot v(t)$ (for calculating the transition rates) of the state vector:
\begin{eqnarray}
\nonumber \gamma(s)&=& sv(s)-v^0 \\
&=&\left[\Lin_0+\Lin_1\mathcal{K}(s)\right]\cdot \left[\mathcal{R}^{-1}(s)-\Lin_1\mathcal{K}(s)\right]^{-1}v^0.  \label{Eq. LT of gamma(t)}
\end{eqnarray}
{The noise affects the system evolution through the operator $\Ker(s)$ as a perturbation to the coherent evolution $\mathcal{R}(s)$. Moreover, the fact that $\Ker(s)$ is a function of $s\mathbb{I}-\Lin_0$ shows that the noise-driven dynamics is mainly determined by the properties of the noise near the system's frequencies.}

With application to actual experiments, Eqs.~\eqref{s-domain} and \eqref{Eq. LT of gamma(t)} link the noise spectrum and the measurement data of some observable of the system, whose expectation value is always a linear function of $v(t)$ or $\gamma(t)$. The  expression of such functional dependence is generally very complicated. However, as will be seen in the next section, it can be simplified under proper selection of the parameters and the system's initial state, so that useful identification formulas can be derived.

\section{Noise Identification in a superconducting charge qubit}\label{Sec:Qubit}

This section will apply the above model to the identification of the noise spectrum in an open quantum system. As a physical example, we study two-level systems implemented by a superconducting charge qubit \cite{Nakamura2000,Duty2004,Astafiev2004,Bladh2005,You2011}. The method developed here can be easily extended to other physical implementations.

\begin{figure}
  \includegraphics[width=3.3in]{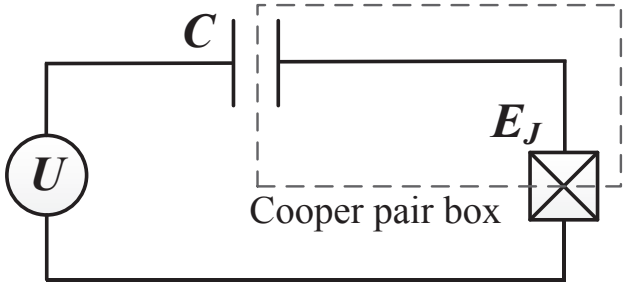}
  \caption{Schematic diagram for the circuit of a superconducting charge qubit. The qubit is encoded by the number of Cooper pairs in the box inside the dashed rectangle (with Josephson energy $E_J$) that is biased by the gate voltage $U$ (proportional to a dimensionless number $n_g$ of charges) through the capacitor (with capacitance $C$). }\label{Circuit}
\end{figure}

{As shown in Fig.~\ref{Circuit}, the superconducting charge qubit is encoded by the charge states $|0\rangle$ and $|1\rangle$ differing by one Cooper pair in the box \cite{Nakamura2000,You2011,Martinis2002,Vion2002,Pashkin2003}. The box is biased by the control gate voltage $U$ via the capacitor $C$. Let $E_J$ be the tunneling energy of the Josephson junction, $E_C$ be the Column energy of the island; and $E_{\rm el}=E_C(1-2n_g)$ be the electrostatic energy that is linear in the dimensionless gate voltage $n_g$ (proportional to $U$ in Fig.~\ref{Circuit}).} Then under the charge state basis $\{|0\rangle,|1\rangle\}$, the qubit's internal Hamiltonian reads
$$H_{\rm tot}=E_J\sigma_x+E_{\rm el}\sigma_z+\sigma_z\otimes C(t),$$
where the bath operator $C(t)$ is responsible for the charge noise (assumed to be dominant in experiments). It is commonly recognized that the noise is mainly induced by fluctuating background charges in the substrate or on the surface \cite{Bergli2006,Yurkevich2010,McDermott2009}, but in the model it can consist of any sources as we are only concerned with the overall correlation properties. {The average charge number, which corresponds to the expectation value of $\sigma_z$ operator, is read out by electrostatically coupling the qubit to a radio-frequency single-electron transistor (SET).}

The density matrix under the Pauli matrix basis
$$\rho(t)=v_0(t)\frac{I_2}{2}+v_1(t)\frac{\sigma_x}{2}+v_2(t)\frac{\sigma_y}{2}+v_3(t)\frac{\sigma_z}{2}$$
is vectorized as a four-dimensional augmented Bloch vector $ v(t)=[v_0(t),v_1(t),v_2(t),v_3(t)]^T$ with $v_0(t)\equiv 1$. Moreover, let $\Delta=\sqrt{E_J^2+E_{\rm el}^2}$ be the energy gap and $\theta  = \arctan (E_J/E_{\rm el})$ be the bias angle, we can rewrite the total Hamiltonian as
\begin{equation}\label{Eq.H0,H1,Coherent}
H_{\rm tot}=\Delta \;\sigma_{\theta }+\sigma_z\otimes C(t),\end{equation}
where $\sigma_\theta =\sigma_z\cos\theta +\sigma_x\sin\theta $. Next, {we assume that the noise correlation time is much shorter than the decay time of the excited state, which validates the Born approximation \cite{Kofman1994,Kofman2004}.} The spectral relationship will be studied between the measurement result and the noise spectrum show under different bias angles and initial states, from which noise identification strategies can be designed.

\subsection{Fermi's Golden Rule}\label{Subsec:FGR}
The Fermi's golden rule has been broadly applied in spectroscopy techniques. In the application to superconducting charge qubit systems \cite{Astafiev2004}, the qubit is initially prepared at the ground state (or the excited state) under some bias voltage, which can be expressed as:
$$|+\rangle=\cos\frac{\theta}{2}|0\rangle+\sin\frac{\theta}{2}|1\rangle,\quad |-\rangle=-\sin\frac{\theta}{2}|0\rangle+\cos\frac{\theta}{2}|1\rangle.$$
The probability for the qubit to stay in the ground state (or the excited state) is measured at sampled time instances, from which the excitation (or decay) rate $\gamma_{\uparrow}(t)$ [or $\gamma_\downarrow(t)$] is measured by interpolating the time-variant curves.

To model this process, it is convenient to express the total Hamiltonian \eqref{Eq.H0,H1,Coherent} in the eigenbasis $\{|+\rangle,|-\rangle\}$, which becomes:
\begin{equation}\label{Eq.Biased basis}
H_{\rm tot}= \Delta\sigma_z+ \sigma_{\theta }\otimes C(t),
\end{equation}
Both $\gamma_{\uparrow}(s)$ or $\gamma_{\downarrow}(s)$ can be calculated from the $z$-component $$\gamma_3(s)=\frac{1}{2}\left[sv_3(s)-v_3^0\right]$$ of $\gamma(s)$, where $v_3^0=1$ for $\gamma_{\uparrow}(s)$ and $v_3^0=-1$ for $\gamma_{\downarrow}(s)$, respectively. Their expressions under any bias angle are given by \eqref{Qs(s)-relaxation} in Appendix \ref{Appendix B}. {Because the noise operator $C(t)$ is weak under the Born approximation, it is reasonable to approximate Eq.~\eqref{Qs(s)-relaxation} by keeping only the lowest-order terms of $\Gamma(s)$ and $\Omega(s)$. This gives}
\begin{eqnarray}\label{Eq.Relaxation Identification}
\label{gup} \gamma_\uparrow(s)&\approx& \frac{\sin^2 \theta }{s}\left[\Gamma_+(s)+\Omega_-(s)\right],\\
\label{gdown}  \gamma_\downarrow(s)&\approx& \frac{\sin^2 \theta }{s}\left[\Gamma_+(s)-\Omega_-(s)\right],
\end{eqnarray}
where $\Gamma_+(s)=\mathscr{L}\left[\Gamma(t)\cos\Delta t\right]$ and $\Omega_-(s)=\mathscr{L}\left[\Omega(t)\sin\Delta t\right]$ are the Laplace transforms of the modulated noise correlation functions with a carrier wave whose frequency is $\Delta$. {Such approximation holds when the noise correlation time is much shorter than the decay time of the excited state.}
One can immediately see that the transition rates from $|+\rangle$ and $|-\rangle$ are differentiated by the noise term $\Omega_-(s)$. {When the time is sufficiently long (far greater than the noise correlation), the transition rate from $|-\rangle$ to $|+\rangle$ approaches a stationary value, which can be calculated by taking the limit $t\rightarrow \infty$ in Eqs.~\eqref{gup} and \eqref{gdown}. In the the frequency domain, these become}
\begin{eqnarray}
\nonumber \bar \gamma_\uparrow & = &\lim_{s\rightarrow 0} s\,\gamma_\uparrow(s)\\
&\approx & {\sin^2 \!\theta }\left[\Gamma_+(0)+\Omega_-(0)\right]=\Phi_{\rm FT}(-\Delta)\sin^2\! \theta,\\
\nonumber \bar \gamma_\downarrow &= &\lim_{s\rightarrow 0} s\,\gamma_\downarrow(s)\\
 &\approx & {\sin^2 \!\theta }\left[\Gamma_+(0)-\Omega_-(0)\right]= \Phi_{\rm FT}(\Delta)\sin^2\! \theta,
\end{eqnarray}
where $\Phi_{\rm FT}(\Delta)$ is the Fourier transform of the correlation function $\Phi(t)$ \footnote{Note that the Fourier transform is a two-sided integral over $\mathbb{R}$, while the Laplace transform is one-sided. It is not difficult to prove that $\Phi_{\rm FT}(\pm \Delta)=\Gamma_+(0)\mp \Omega_-(0)$, i.e., the two-sided Fourier transformed function $\Phi_{\rm FT}(\pm \Delta)$ is decomposed into two one-sided integrals.}. This exactly recovers the formulas adopted in \cite{Astafiev2004,Almog2011} according to the Fermi's golden rule, which holds when the measurement time $t$ is much smaller than the inverse of the stationary decay rate, and far greater than the autocorrelation time of the noise.

The Fermi's golden rule shows that the spectral density of a noise at some frequency $\Delta$ is proportional to the transition rate of a resonating two-level system. Thus, a transition-frequency tunable qubit can be used as a spectrum analyzer. In the charge qubit system, the transition frequency $\Delta=E_J\sin^{-1}\theta$ is tuned by the bias angle $\theta$. {The identification scheme based on the Fermi's golden rule is operationally convenient, but the precision may be limited due to the use of perturbation and asymptotic approximations, which equivalently requires that the system behaves approximately Markovian at the time of measurement (i.e., the environment has no memory effects on the system's evolution).

\subsection{Noise identification at the optimal point}\label{Subsec:OptPoint}
The identification formula based on the Fermi's golden rule is obtained by measuring the decay rate under various bias angles $\theta$. Actually, it is possible to find better identification schemes from the full measurement data under a fixed bias angle. For example, the identification formula Eq.~\eqref{Qs(s)-relaxation} can be greatly simplified at $\theta=\pi/2$ (i.e., the so-called optimal point in superconducting qubit systems) as below
\begin{equation}\label{relaxation charge}
\gamma_\uparrow(s) = \frac{\Gamma_+(s)+\Omega_-(s)}{\Gamma_+(s)+s},~\gamma_\downarrow(s) = \frac{\Gamma_+(s)-\Omega_-(s)}{\Gamma_+(s)+s}.
\end{equation}
The transition rates differ only by $\Omega_-(s)$ in the numerators, which has its physical origin from the non-uniform thermal distribution of populations on the qubit levels. Eq.~\eqref{relaxation charge} can be easily reversed to obtain an identification formula:
\begin{eqnarray}\label{Eq.Relaxation Identification a=90}
\Gamma_+(s)&=& \frac{s\left[\gamma_\uparrow(s)+\gamma_\downarrow(s)\right]}{2-\left[\gamma_\uparrow(s)+\gamma_\downarrow(s)\right]},\\
\Omega_-(s)&=&\frac{s\left[\gamma_\uparrow(s)-\gamma_\downarrow(s)\right]}{2-\left[\gamma_\uparrow(s)-\gamma_\downarrow(s)\right]},
\end{eqnarray}
for $\Gamma_+(s)$ and $\Omega_-(s)$, respectively, corresponding to the modulated signals $\Gamma(t)\cos\Delta t$ and $\Omega(t)\sin \Delta t$, respectively. Because no asymptotic approximation is taken here, the identification is expected to be more precise and {more economic because only measurement data at the optimal point are required}.

The above formula can be used to identify the modulated signals $\Gamma(t)\cos\Delta t$ [or $\Omega(t)\sin \Delta t$]. However, from them one cannot obtain the full correlation function {because the values of $\Gamma(t)$ [or $\Omega(t)$] at the time instances $t=\left(k+\frac{1}{2}\right)\frac{\pi}{\Delta}$ (or $t=\frac{k\pi}{\Delta}$), where $k$ is an arbitrary nonnegative integer, cannot be determined.} Also, the presence of high-frequency measurement noise may cause large error in the calculation of $\gamma_\uparrow(t)$ or $\gamma_\downarrow(t)$. Nevertheless, this idea can be extended to the following different experimental setup \cite{Nakamura2000}, with which at least the spectrum of $\Gamma(t)$ can be identified at the optimal point.

As shown in Fig.~\ref{Coherent}(a), the qubit is prepared at the zero-charge state $|0\rangle$ and released for free evolution under some bias angle $\theta$. The average charge number is then measured at sampled time instants [see Fig.\ref{Coherent}(b)], from which coherent oscillations can be observed \cite{Nakamura2000}. The presence of charge noise $C(t)$ will damp the oscillation via.

\begin{figure}
  \includegraphics[width=3in]{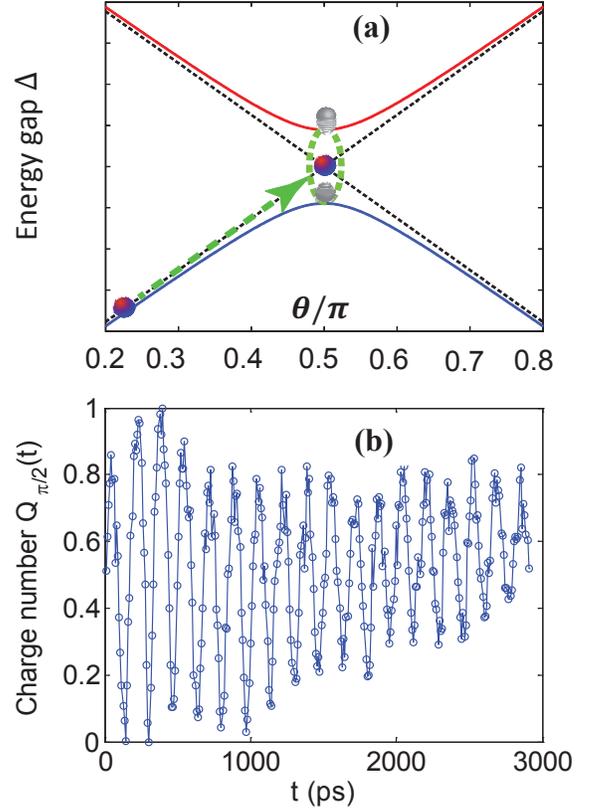}
  \caption{(color online) Schematic diagram for identifying the noise $\Gamma(t)$. (a) The energy gap versus the bias angle, where both quantities are defined below Eq.~\eqref{Eq.Biased basis}. The charge qubit is initially prepared in the zero-charge state, corresponding to the ground state far from the optimal point. Then, a gate pulse is applied to induce coherent oscillations at some bias angle $\theta $ (e.g., the optimal point $\theta =\frac{\pi}{2}$), which is determined by the height of the pulse. The pulse persists for a time $t$, after which the average charge number is measured. (b) The coherent oscillation curve (with data collected from NEC and RIKEN) of the average charge number $Q_{\pi/2}(t)$ versus the time $t$, which is measured at the optimal point $\theta=\pi/2$.}\label{Coherent}
\end{figure}

Because the average charge number is measured, it is more convenient to use the original total Hamiltonian \eqref{Eq.Biased basis} in the charge states basis $\{|0\rangle,~|1\rangle\}$. Correspondingly, the average charge number $Q_\theta(t)$ is calculated from the $z$-component of the Bloch vector:
\begin{equation}
Q_\theta(t)= \frac{1}{2}\left[1-{v_3(t)}\right]~\rightarrow~Q_\theta(s)= \frac{1}{2}\left[\frac{1}{s}-{v_3(s)}\right].
\end{equation}
In Appendix \ref{Appendix C}, the functional dependence of $Q_\theta(t)$ on the noise spectrum is derived [see Eq.~\eqref{Qs(s)-dephasing}]. At the optimal point $\theta =\frac{\pi}{2}$ (corresponding to $E_{\rm el}=0$), the formula becomes
\begin{equation}\label{Optimal Point}
Q_{{\pi}/{2}}( s)=\frac{ \Delta^2}{2s\left[s^2+s  \Gamma( s) + \Delta^2\right]},
\end{equation}
where only $\Gamma(s)$ is present but $\Omega(s)$ disappears. This can be used to separate the identification of $\Gamma(t)$ and $\Omega(t)$. By reversing Eq.~\eqref{Optimal Point} at $s=i\w$, we obtain an identification formula for $\Gamma(t)$
\begin{equation}\label{Identification 1}
\Gamma(i\w)=-\frac{\Delta^2}{2\w^2Q_{{\pi}/{2}}(i\w)}-i\left(\w-\frac{\w_0^2}{\w}\right),
\end{equation}
where $\w_0=E_J/\hbar$; and $Q_{{\pi}/{2}}(i\w)$ is obtained by Laplace transforming the measured coherent oscillation curve in the time domain. From \eqref{Identification 1}, the Fourier transform of the noise spectrum, which is more often used in practice, can be obtained as
\begin{equation}\label{Identification 1'}
\Gamma_{\rm FT}(\w)=2~{\rm Re}\Gamma(i\omega)=-\w_0^2{\rm Re}\left[\frac{1}{\w^2Q_{{\pi}/{2}}(i\w)}\right],
\end{equation}
where the symmetry property $\Gamma(-t)=\Gamma(t)$ of the noise $\Gamma(t)$ was utilized [see \eqref{Eq:symmetry of Phi} in Appendix \ref{Appendix A}].

Since there is an $\w^2$ term in the denominator of Eq.~\eqref{Identification 1'}, which may cause significant numerical error in the low-frequency regime, we can expressing the noise spectrum as a function of the AC component $Q^{\rm AC}_{\pi/2}(t)=Q_{\pi/2}(t)-0.5$ of the measurement data $Q_{\pi/2}(t)$, which leads to the following identification formula
\begin{equation}\label{Identification 2}
\Gamma_{\rm FT}(\w)=-\w_0^2{\rm Re}\,\left[\frac{2Q^{\rm AC}_{\pi/2}(i\w)}{0.5+i\w  Q^{\rm AC}_{\pi/2}(i\w)}\right]
\end{equation}
from Eq.~\eqref{Optimal Point}.

The above formulas suggest that an identification scheme can be devised as shown in Fig.\ref{QubitID}. For illustration, the formula \eqref{Identification 1} is used to identify the noise $\Gamma(t)$ from the experimental data collected in NEC and RIKEN [see Fig.~\ref{Coherent}(b)] for testing coherent oscillations at the optimal point. The average charge number was measured with a time delay sweeping over a time period of $T=2900$~ps with sampling time $\delta t=9$~ps. From the Laplace transform (at $s=i\w$) of the oscillation curve, the characteristic frequency is read as $\Delta\approx 6.0$~GHz from the peak in Fig.~\ref{IDRESULT}(a). Then the spectrum of $\Gamma(t)$ was calculated Eq.~\eqref{Identification 2} as shown in Fig.~\ref{IDRESULT}(b).

\begin{figure}
  \includegraphics[width=3in]{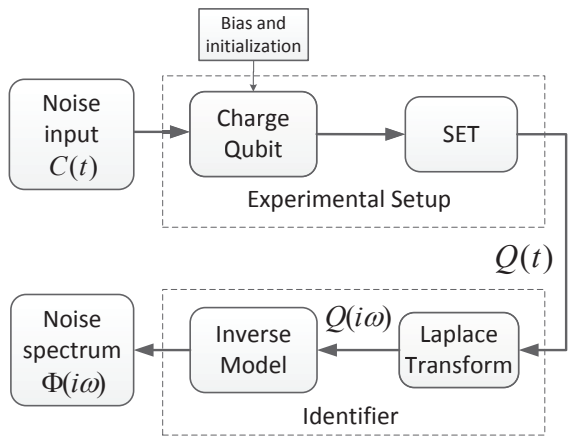}
  \caption{Schematic diagram for noise identification in a superconducting charge qubit. The average charge number $Q(t)$ is measured and transformed into the frequency domain, and then used to calculate the noise spectrum via identification formulas \eqref{Identification 1'} or \eqref{Identification 2} at the optimal point.}\label{QubitID}
\end{figure}
This example illustrates how the identification strategy works with a simple experiment design without using additional control pulses or devices. The identification only uses the measurement data at the optimal point, but the samples should cover a sufficiently broad time interval. The identification error could come from many possible factors, including the measurement noise (not the noise coupled to the qubit) from the SET device. Moreover, the data collected from the experiment has a finite number of data points and finite sampling time period, which restricts the precision and frequency regime of the identification results. To improve the identification over a wider range, the data should be more densely sampled over a sufficiently long time period until the oscillation decays to approximately zero.

{Moreover, the linewidth of the coherent oscillation spectrum shown in Fig.~\ref{QubitID}(a) is comparable to the width of $\Gamma_{\rm FT}(\w)$ in Fig.~\ref{QubitID}(b), which seems to imply that the noise correlation time is comparable to the decay time. This could be due to the strong noise coupling (where the Born approximation, as well as the above derived identification formulas, are not valid), but it could also be from measurement errors (including the measurement noise and the finite number of data points). These factors should be further analyzed.}

\begin{figure}
  \includegraphics[width=3.5in]{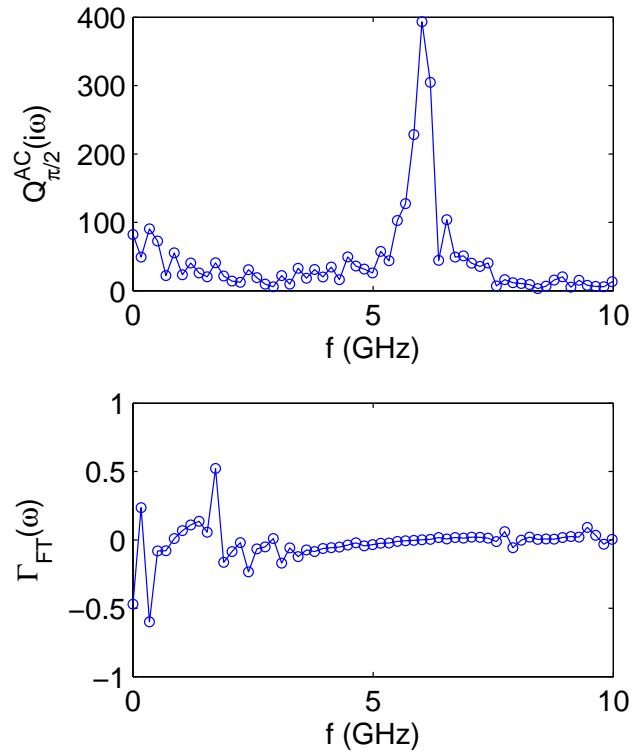}\\
  \caption{(color online) (a) The {absolute value of} the Laplace transform of $Q^{\rm AC}_{\pi/2}(t)=Q_{\pi/2}(t)-0.5$, where $Q_{\pi/2}(t)$ is the average charge number at the optimal point. (b) The identified noise spectrum $\Gamma_{\rm FT}(\omega)$ at the optimal point.}\label{IDRESULT}
\end{figure}


\section{Conclusion}\label{Sec:Conclusion}
We have presented a noise-identification approach that can improve the precision based on a frequency-domain model for the system's non-Markovian dynamics {with less amount of measurement data.} With applications to qubit systems, the Fermi's golden rule can be naturally derived under proper approximations from this model. The advantage of this model is that a simple identification strategy can be obtained without aid of additional control pulses or devices. The obtained identification formula at the optimal point is illustrated with applications to superconducting qubits, which requires only measurements at the optimal point.

It was also seen that the quality of the identification is limited due to the finite data points and possibly severe measurement noises in the experiment. Moreover, the efficient identification of the noise term $\Omega(t)$, which is non-negligible under ultra-low temperatures, is to be studied. In principle, the full expression \eqref{Qs(s)-dephasing} derived in this paper can be used to extract the noise spectrum of $\Omega(t)$ with data obtained under biased configurations. However, sophisticated numerical algorithms are to be designed due to the complexity of the formulas. It is also possible to design other measurement schemes with which the identification formulation can be simplified. These problems are to be explored in the future.

\begin{acknowledgements}
RBW, TFL, JZ and YXL are supported by the Tsinghua National Laboratory for Information Science and Technology (TNList) Cross-discipline Foundation and NSFC under Grant
Nos. 60904034, 61106121, 61174084, 61134008, 10975080, 61025022 and 60836001. Yu.A.P. acknowledges partial support by the Royal Society and Wolfson Foundation. FN is partially supported by the ARO, JSPS-RFBR contract No. 12-02-92100, Grant-in-Aid for Scientific Research (S), MEXT Kakenhi on Quantum Cybernetics, and the JSPS via its FIRST program.
\end{acknowledgements}
\appendix
\section{Properties of the noise correlation function}\label{Appendix A}
According to the definition \eqref{Eq: def of Phi} of
the noise correlation function, it is easy to examine the following symmetry properties:
\begin{equation}\label{Eq:symmetry of Phi}
    \Gamma(-\tau)=\Gamma(\tau),\quad \Omega(-\tau)=-\Omega(\tau),\quad \Phi(-\tau)=\Phi^*(\tau).
\end{equation}
Owing to these symmetry properties, the Fourier transform of the symmetric function $\Gamma(\tau)$ is a purely real-number valued function, while the Fourier transform of the antisymmetric function $\Omega(\tau)$ is purely imaginary.

In this paper, both the Laplace and Fourier transforms are involved to describe the noise spectrum. The major difference between them is that the Fourier transform is a two-sided integral in the time domain from $\tau=-\infty$ to $\tau=\infty$, while the Laplace transform is one-sided (i.e., from $\tau=0^-$ to $\tau=\infty$). Because $\Phi(\tau)$ is evaluated for both negative and positive times, the noise spectrum is generally defined as the Fourier transform of the correlation function $\Phi(\tau)$ \cite{Schoelkopf2003,Clerk2010}. However, in the frequency-domain model \eqref{Eq. LT of gamma(t)} derived from the generalized master equation \eqref{rho-ME}, the Laplace transform has to be adopted because only the positive time-correlation function is involved. Another reason for using the Laplace transform is that the system is always prepared at some initial state in any realistic measurement, regarding to which the Fourier transform is physically not applicable. Nevertheless, owing to the symmetry properties, it is sufficient to use the positive-time branch, from which we can recover full noise correlation function.

Therefore, the Laplace transform will be adopted corresponding to the positive time-correlation function. To avoid confusion, the Fourier transform of a time-variant function, say $\Phi(t)$, will be denoted by $\Phi_{\rm FT}(\omega)$ with a subscript. $\Phi_{\rm FT}(\omega)$ is generally different from the Laplace transform $\Phi(i\omega)$ evaluated at $s=i\omega$, but the symmetry property of the correlation functions guarantees the following relations:
\begin{equation}\label{Eq:LT and FT}
    \Gamma_{\rm FT}(\omega)=2{\rm Re}\Gamma(i\omega),~~     \Omega_{\rm FT}(\omega)=2{\rm Im}\Omega(i\omega),
\end{equation}
which will be used to extract the temporal correlation function from the noise spectrum obtained via the inverse Fourier transform.

\section{Frequency domain Derivations in Sec. \ref{Sec:Qubit}A}\label{Appendix B}
Firstly, suppose that the eigenvalues of $\Lin_0$ are $x_1,\cdots,x_n$, and that $\Lin_0$ is diagonalized by $P$, then
the matrix $s$-function $\Gamma(s\mathbb{I}-\Lin_0)$ can be calculated as follows:
$$\Gamma(s\mathbb{I}-\Lin_0)=P^{-1}{\rm diag}\{\Gamma(s-x_1),\cdots,\Gamma(s-x_n)\}P,$$
and in a similar way $\Omega(s\mathbb{I}-\Lin_0)$ can be evaluated.

With respect to the Hamiltonian \eqref{Eq.Biased basis}, the system is initially prepared at the excited state corresponding to the Bloch vector $v^0=[1,0,0,-1]^T$. Correspondingly, the matrix representation $\Lin_0$ of $\Delta \sigma_z $ is diagonalized as
$$\Lin_0 =P^{-1}{\rm diag}\{0,0,i\Delta,-i\Delta\}P,$$
where
\begin{equation}\label{Eq-P alpha}
P =\left(
\begin{array}{cccc}
 1 & 0  & 0 & 0 \\
 0 & 1  & 1 & 0 \\
 0 & -i & i & 0 \\
 0 &  0 & 0 & 1
\end{array}
\right).
\end{equation}
and hence
$$\Gamma(sI-\Lin_0 )=P^{-1}{\rm diag}\{\Gamma(s),\Gamma(s),\Gamma(s-i\Delta),\Gamma(s+i\Delta)\}P.$$

When the qubit is released from the excited state, from Eq.~\eqref{s-domain}, the corresponding transition rate under the bias angle $\theta $ can be expressed as
\begin{equation}\label{Qs(s)-relaxation}
\gamma_\downarrow ( s)=\sin^2\theta\cdot\frac{N_0(s)\cot^2\theta+N_1(s)}{ D_0( s)\cot^2\theta + D_1( s)},
\end{equation}
where, by dropping the argument ``s" for simplicity, we have
\begin{eqnarray*}  N_0( s) &=& (s^2+\Delta^2)\Gamma_++s(\Gamma_+^2+\Gamma_-^2)\\
&& (\Omega- \Omega_+)(\Delta\Gamma_+-s\Gamma_-)-\Omega_-(s^2+s\Gamma_++\Delta^2)\\
  N_1(s) &=& (s^2+s\Gamma + \Delta^2) (\Gamma_+ - \Omega_-)\\
  D_0( s) &=&s \left[(s + \Gamma_+)^2 + (\Gamma_- + \Delta)^2\right]\\
  D_1( s) &=& (s + \Gamma_+) (s^2+ s \Gamma  + \Delta^2).
\end{eqnarray*}
Here, we use the following notations:
\begin{eqnarray}
\nonumber {\Gamma}_+(s)&=& \frac{{\Gamma}(s+i\Delta)+{\Gamma}(s-i\Delta)}{2}\\
&=&\mathscr{L}\left[ \xi(t)\cos \Delta t\right], \label{Eq.XI+}\\
\nonumber {\Gamma}_-(s)&=& \frac{{\Gamma}(s+i\Delta)-{\Gamma}(s-i\Delta)}{2i}\\
&=&\mathscr{L}\left[\gamma(t)\sin \Delta t\right], \label{Eq.XI-}
\end{eqnarray}
and in the same way are $\Omega_\pm(s)$ defined. These represent signals modulated by sinusoidal waves with the carrier frequency $\Delta$. Such noise spectral functions shifted by the oscillation frequency $\pm \Delta$ exhibit the interplay between the qubit and its environments.

When the qubit is prepared in the ground state, the expression of $\gamma_\uparrow(s)$ is similar to Eq.~\eqref{Qs(s)-relaxation} except that the signs of all terms of $\Omega(s)$ and $\Omega_{\pm}(s)$ are flipped. This shows that the presence of $\Omega(t)$ in the noise correlation function causes the difference between the static average charge number and the corresponding transition rates.

\section{Frequency domain Derivations in Sec. \ref{Sec:Qubit}B}\label{Appendix C}
The extended Bloch vector corresponding to the initial zero-charge state with Bloch vector being $v^0=[1,0,0,1]^T$. We first diagonalize $\Lin_0 =\Delta\sigma_\theta$ as
$$\Lin_0 =P_\theta ^{-1}{\rm diag}\{0,0,i\Delta,-i\Delta\}P_\theta ,$$
where
\begin{equation}\label{Eq-P alpha}
P_\theta =\left(
\begin{array}{cccc}
 1 & 0 & 0 & 0 \\
 0 & \tan\theta  & -\cot \theta  & -\cot\theta  \\
 0 & 0 & i \csc\theta  & -i \csc\theta  \\
 0 & 1 & 1 & 1
\end{array}
\right).
\end{equation}
Consequently, we have
$$\Gamma(sI-\Lin_0 )=P_\theta ^{-1}{\rm diag}\{\Gamma(s),\Gamma(s),\Gamma(s-i\Delta),\Gamma(s+i\Delta)\}P_\theta ,$$
and from Eq.~\eqref{s-domain}, the average charge $Q_\theta (t)$ for the bias angle $\theta $ can be expressed as
\begin{equation}\label{Qs(s)-dephasing}
Q_\theta ( s)=\frac{\Delta}{2 s}\cdot\frac{N_0( s)\cos\theta + N_1( s)}{ D_0( s)\;\cot^2\theta+ D_1( s)},
\end{equation}
where
\begin{eqnarray*}
 N_0( s) &=& (s+\Gamma_+)\left[ \Omega_+- \Omega ( s)\right]+ (\Delta+\Gamma_-)  \Omega_- , \\
  N_1( s)  &=& \Delta(s+\Gamma_+) ,  \\
  D_0( s)  &=&  s \left[ (s+\Gamma_+) ^2+(\Delta+\Gamma_-) ^2\right], \\
  D_1( s) &=&  (s+\Gamma_+) \left( s^2+s  \Gamma + \Delta^2\right). \end{eqnarray*}


\end{document}